# Toward a Local Perspective on Online Collaboration

HANI SAFADI and SAMER FARAJ, Desautels Faculty of Management, McGill University

## 1. INTRODUCTION

The position of a node within a complex network is closely related to the function of that node. For example, in gene interaction networks, the connectivity of a gene and its rate of evolution are interdependent [Fraser et al. 2002]. This observation extends to networks formed via social interaction. The position of actors within social networks, both online and offline, influences their social outcomes such as knowledge distribution, wages, job placement, promotion opportunities, creativity, innovation, political success, social support, productivity, and performance [Aral and Van Alstyne 2011].

In this paper, we focus on how the position/embeddedness of members in an online community of innovation relates to their knowledge contribution to the community. We depart from previous research by bringing a different perspective to examine position within a network: we focus on the *local position* of members rather than the traditionally considered *global position*. This differentiation has significant theoretical and methodological implications. On the theoretical front, a local view of position implies a more confined and local organization of work in online communities than previously thought. From a methodological perspective, evaluating the local structure of large networks involves radically different algorithms that have only recently become feasible with the increase of processing power.

## 2. KNOWLEDGE CREATION IN ONLINE COMMUNITIES

The process of innovation in many organizations relies on the availability of local expertise in in-house research and development teams. Online communities of innovation employ different organizational structures enlisting contributions from a large number of participants. For example, in innovation contests, a problem is broadcasted to many peripheral solvers who compete on providing a solution to earn a prize [Lakhani 2006]. Another example is open-source software development by communities of spatially, temporally, and organizationally distributed programmers [Cummings et al. 2009].

We study a community of innovation centered on an open-source Electronic Medical Record (EMR) named OSCAR [OSCAR 2012]. This community is primarily Canadian, has been in existence for a decade, and has developed an EMR that is rapidly diffusing (currently used by over 1,500 Canadian doctors to follow over a million patients). OSCAR EMR is freely available and is gaining market share against commercial products typically costing $25,000 per year per user. Contrary to other open-source projects, OSCAR is embedded in a community of doctors rather than a community of programmers.

Within the OSCAR community, we examine how members' positions relate to their knowledge contribution. Knowledge is an important resource in organizations [Grant 1996]. Moreover, knowledge creation is the goal of many online communities (e.g. wikis, forums and open-source projects). We focus our attention to individual knowledge contribution within the process of knowledge creation. More specifically, we examine members' contribution of small chunks of code in the discussion forums. This allows us to study contribution and the member-level and relate it to position within the community.

## 3. NEW PERSPECTIVE TO EXAMINE POSITION

Early research in social networks examined how an actor's influence in a network depends on the quality of his/her immediate connections with others. Strong ties and frequent interactions among team members lead them to build mutual understanding of their expertise and achieve better performance [Wegner et al. 1985]. On the other hand, weak ties among actors are important for reaching opportunities beyond local communities that are not accessible via strong ties [Granovetter 1973]. Combining the two findings, a great influence can be gained by bridging a gap between and establishing strong ties with weakly tied actors having complementary resources [Burt 1992].

Later research examined how the global position within a social networks influences social outcomes. First, there is a consensus on the important role of central members in driving innovation [Dahlander and Frederiksen 2011]. In online communities, central members have more social capital and contribute more knowledge and provide better answers to members' requests [Wasko and Faraj 2005]. Second, members who are not central but span boundaries between subgroups in their communities or between different communities are capable of bringing fresh ideas and solutions to problems that central members [Jeppesen and Lakhani 2010]. This leads to a *core-periphery* model in which the role of an actor corresponds to his/her global position in the network [Berdou 2011, p. 12].

### 3.1 Toward a Local View of Structure

Recent research on large scale collaboration began to challenge the global perspective on organizing work in online communities and open-source development. For example, it is been suggested that in many open-source projects, the majority of contributions comes from *lonely* developers who are working in *caves* and seldom interact with others [Krishnamurthy 2002]. This is supported by empirical evidence from repositories such as SourceForge where the larger a project is, the smaller the number of its administrators are. In addition, larger open-source development teams tend to have flatter hierarchies and more decentralized communication with uneven degree of participation [Crowston and Howison 2005]. Finally, while the organizational structure in traditional organizations is preestablished and visible to all actors, it is emergent and temporal in online communities [Butler et al. 2008].

We propose that *the local position of a member in an online community is much important in driving contribution to the community than his/her global position*. For example, Figure 1 depicts a hypothetical online community of innovation. We argue that we can learn more about the knowledge contribution of the starred member by examining her relationships within her neighborhood (the smaller circle) than by examining her relationships with every other member in the community.

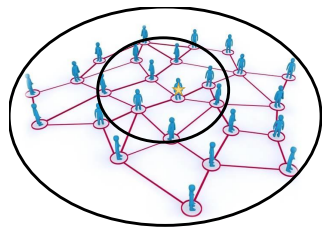

**Fig. 1:** Local (within the small circle) and global position (within the big circle) of the starred member

### 4. METHODS

We examine three global measurements of a member's position in the community: *centrality*, *boundary spanning* and location within a core-periphery hierarchy or *coreness*. We operationalize centrality using closeness centrality which measures how long it takes to sequentially disseminate a message from a node to all other nodes in the network [Brandes and Pich 2007]. To operationalize boundary spanning, we use betweeness centrality which measures how a node mediates between shortest paths among all node pairs in the network [Brandes 2001]. Finally, we operationalize the coreness of a node using the k-core number which measures the membership of a node within a cohesive subnetwork that requires at least k edge removals in order to disconnect [Seidman 1983]. Note that The operationalization of each global measurement requires complete network information to compute.

To measure a member's local position, we adapt a type of local features first introduced for studying structure in biological networks called *graphlets* (Figure 2). Graphlets are small connected non-isomorphic induced subgraphs of the large network [Milenkoviæ and Pržulj 2008]. Graphlets can be classified according to their size. We consider here 2-node, 3-node, and 4-node graphlets with one, two and six graphlets in each respectively. Similar to how we can compare two nodes based on their centrality or boundary spanning, we can compare two nodes based on the graphlets that happen to include them. An *orbit* is defined to be the location where a node touches a graphlet [Pržulj 2007].

Orbits can be classified depending on their local centrality. There are orbits that touch one edge (0, 1, 4, 6, and 9), orbits that touch two edges (3, 8, 10, and 12) and orbits touching three edges such (7, 11, 13, and 14). Orbits can also be classified depending on their local boundary spanning by counting the number of disjoint components the graphlet breaks to when deleting the orbit: zero (0, 1, 3, 4, 6, 8, 9, 10, 12, 13, 14), one (2, 5, 11) and two (7). Note that while assessing global properties of structure requires the consideration of all members and their relationships in the community, the assessment of local properties requires information about a limited number of members and their relationships.

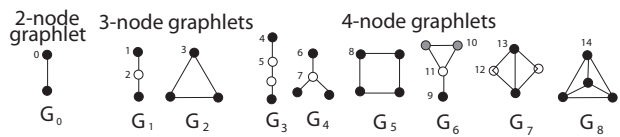

**Fig. 2:** The 9 graphlets of 3 and 4 and their associated orbits (adapted from [Pržulj 2007])

We regress members' codified knowledge contribution on their global properties (centrality, spanning and coreness), local properties (local centrality and local boundary spanning) and personal characteristics control (tenure and profession) using a log-log model. This model suits the setting better than an absolute one because structural features are better thought as moderators of contribution than absolute predictors. In order to determine the relative importance of global and local measurements, we run three separate regressions: (1) global measurements (2) local measurements, and (3) both local and global measurements. We compare the three regressions with two model difference tests.

## 5. PRELIMINARY RESULTS

The regression analysis shows that local position is a better predictor of knowledge contribution than global position. Local measurements alone explain 85% of the dependent variable.

## 6. DISCUSSION AND CONCLUSION

This study seeks to deepen our understanding of the structural properties of large scale collaboration in online communities of innovation and the role that position in the community plays in determining knowledge contribution. Contrary to previous research, we argue for a more local perspective when examining online collaboration. We demonstrate that a member's centrality and spanning within his/her local neighborhood is a better predictor of contribution than global centrality and spanning within the whole community. Methodologically, we leverage recent research in network science and develop a comprehensive apparatus to examine position on both the macro and the micro scales.

REFERENCES

S. Aral and M. Van Alstyne. 2011. The Diversity-Bandwidth Trade-off. *Amer. J. Sociology* 117, 1 (2011), 90–171.

E. Berdou. 2011. *Organization in Open Source Communities: At the Crossroads of the Gift and Market Economies*. Vol. 15. Taylor & Francis US.

U. Brandes. 2001. A faster algorithm for betweenness centrality*. *Journal of Mathematical Sociology* 25, 2 (2001), 163–177.

U. Brandes and C. Pich. 2007. Centrality estimation in large networks. *International Journal of Bifurcation and Chaos* 17, 07 (2007), 2303–2318.

R. Burt. 1992. *Structural holes: The social structure of competition*. Harvard University Press.

B. Butler, E. Joyce, and J. Pike. 2008. Don't look now, but we've created a bureaucracy: the nature and roles of policies and rules in wikipedia. In *Proceedings of the twenty-sixth annual SIGCHI conference on Human factors in computing systems*. ACM, 1357227, 1101–1110.

K. Crowston and J. Howison. 2005. The social structure of free and open source software development. *First Monday* 10, 2-7 (2005).

J. Cummings, E. A., and P. C.K. 2009. Crossing Spatial and Temporal Boundaries in Globally Distributed Projects: A Relational Model of Coordination Delay. *Information Systems Research* 20, 3 (2009), 420–439.

L. Dahlander and L. Frederiksen. 2011. The core and cosmopolitans: A relational view of innovation in user communities. *Organization science* (2011).

H. B. Fraser, A. E. Hirsh, L. M. Steinmetz, C. Scharfe, and M. W. Feldman. 2002. Evolutionary rate in the protein interaction network. *Science Signalling* 296, 5568 (2002), 750.

M. S. Granovetter. 1973. The strength of weak ties. *American journal of sociology* (1973), 1360–1380.

R. Grant. 1996. Toward a knowledge-based theory of the firm. *Strategic management journal* 17 (1996), 109–122.

L. Jeppesen and K. Lakhani. 2010. Marginality and problem-solving effectiveness in broadcast search. *Organization science* 21, 5 (2010), 1016–1033.

S. Krishnamurthy. 2002. Cave or community?: An empirical examination of 100 mature open source projects. *First Monday* (2002).

K. R. Lakhani. 2006. Broadcast Search in Problem Solving: Attracting Solutions from the Periphery. In *Technology Management for the Global Future, 2006. PICMET 2006*, Vol. 6. IEEE, 2450–2468.

T. Milenkoviæ and N. Pržulj. 2008. Uncovering biological network function via graphlet degree signatures. *Cancer informatics* 6 (2008), 257.

OSCAR. 2012. Oscar community web site. (2012). http://www.oscarcanada.org/

N. Pržulj. 2007. Biological network comparison using graphlet degree distribution. *Bioinformatics* 23, 2 (2007), e177–e183.

S. B. Seidman. 1983. Network structure and minimum degree. *Social networks* 5, 3 (1983), 269–287.

M. Wasko and S. Faraj. 2005. Why Should I Share: Examining Social Capital and Knowledge Contribution in Electronic Networks of Practice. *Management Information Systems Quarterly* 29, 1 (2005), 35–47.

D. M. Wegner, T. Giuliano, and P. T. Hertel. 1985. Cognitive interdependence in close relationships. In *Compatible and incompatible relationships*. Springer, 253–276.